\RequirePackage{fix-cm}
\documentclass[twocolumn]{svjour3}          % twocolumn
\smartqed  % flush right qed marks, e.g. at end of proof
\usepackage{graphicx}
\usepackage[flushleft]{threeparttable}
\usepackage{csquotes}
\usepackage[hyphens]{url}
\usepackage{multirow}
\usepackage[tikz]{bclogo}
\usepackage{csquotes}
\usepackage[justification=centering]{caption}
\usepackage[T1]{fontenc}
\usepackage{placeins}

\usepackage{float}
\usepackage[compact]{titlesec}
\begin{document}
\sloppy
\title{How Much Data Analytics is Enough? \subtitle {The ROI of Machine Learning Classification and its Application to Requirements Dependency Classification} %\thanks{Grants or other notes
%about the article that should go on the front page should be
%placed here. General acknowledgments should be placed at the end of the article.}
}
% \subtitle{Do you have a subtitle?\\ If so, write it here}

%\titlerunning{Short form of title}        % if too long for running head

\author{*Gouri Deshpande         \and
        Guenther Ruhe \and Chad Saunders
}

%\authorrunning{Short form of author list} % if too long for running head

\institute{Gouri Deshpande \at
              Dept. Computer Science\\
                University of Calgary
            %   Tel.: +123-45-678910\\
            %   Fax: +123-45-678910\\
            \\
            \email{gouri.deshpande@ucalgary.ca}           %  \\
%             \emph{Present address:} of F. Author  %  if needed
           \and
          Guenther Ruhe \at
              Dept. Computer Science\\
              Dept. Electrical and Software Engineering\\
                University of Calgary  \\
            \email{ruhe@ucalgary.ca}
            \and
          Chad Saunders \at
              Haskayne School of Business\\
                University of Calgary  \\
            \email{wsaunder@ucalgary.ca}
}

\date{Received: date / Accepted: date}
% The correct dates will be entered by the editor

\maketitle

\begin{abstract}
Machine Learning (ML) can substantially improve the efficiency and effectiveness of organizations and is widely used for different purposes within  Software Engineering.  However, the selection and implementation of ML techniques rely almost exclusively on accuracy criteria.  Thus, for organizations wishing to realize the benefits of ML investments, this narrow approach ignores crucial considerations around the anticipated costs of the ML activities across the ML life-cycle, while failing to account for the benefits that are likely to accrue from the proposed activity.  We present findings for an approach that addresses this gap by enhancing the accuracy criterion with return on investment (ROI) considerations.  Specifically, we analyze the performance of the two state-of-the-art ML techniques: Random Forest and Bidirectional Encoder Representations from Transformers (BERT), based on accuracy and ROI for two publicly available data sets. Specifically, we compare decision-making on requirements dependency extraction (i) exclusively based on accuracy and (ii) extended to include ROI analysis. As a result, we propose recommendations for selecting ML classification techniques based on the degree of training data used. Our findings indicate that considering ROI as additional criteria can drastically influence ML selection when compared to decisions based on accuracy as the sole criterion.

% \PACS{PACS code1 \and PACS code2 \and more}
% \subclass{MSC code1 \and MSC code2 \and more}
\end{abstract}

\section{Introduction}
%Vast amount of of data is produced by software-intensive systems every day. As a result Software Analytics also called as Data Analytics is used extensively to gain insights from data.  As existing tools are typically quite generic and can often not provide the desired depth of product-specific and stakeholder-targeted information, there is often a need for customized softwar analytics or business intelligence (DA/BI) solutions that leverage the full potential of modern machine learning (ML) techniques.  
Machine Learning (ML) includes methods, tools, and techniques for inferring models from data and has provided successful applications of classification and prediction algorithms. In the area of software development and evolution, a recent study \cite{shafiq2020machine} revealed that there is a spectrum of applications of ML across the software development life-cycle, with most of the applications belonging to the category of \textit{Quality Assurance and Analytics}. 

There exists an extensive variety of ML algorithms and this pool is growing steadily. A recent study \cite{shafiq2020machine} listed Decision Trees, Naive Bayes, and Random Forrest as the techniques most frequently applied in Software Engineering. However, it is important to determine which algorithm works well for a given problem and which are less effective. The performance of any ML technique is generally measured in terms of accuracy (or similar measures). However, the success of ML does not only depend on the algorithms used because ML is a process with various interdependent steps and the investments made in this process need to be related to the return gained from its results. This paper puts estimating the return-on-investment (ROI) of ML in the spotlight. ROI  is most widely used in the context of business analysis, which we extend to ML classification problems. In particular, we focus on the decision-making of ML method selection, i.e., to determine when to stop the process and how much additional investment is needed to achieve a target goal (result).

The most important prerequisite for generating accurate ML models is high-quality training data, however securing such data is often an arduous task. Additionally, engineering and selecting appropriate features is especially time-consuming and requires a vast amount of effort and resources \cite{figalist2020end}. The benefits gained from the application of ML can be dramatically offset due to data collection and data pre-processing activities, which incur substantial costs and effort.

ROI is of great interest in engineering and business, where it is widely used as a guide for decision-making. This is true in Software Engineering (SE) as well. For example, Silverio et al. \cite{martinez2013rearm} evaluated cost-benefit analysis for the adoption of software reference architectures for optimizing architectural decision-making. Cleland et al. \cite{cleland2004heterogeneous} studied the ROI of heterogeneous solutions for the improvement of requirements traceability. However, the recent data explosion in the form of big data and advances in Machine Learning (ML) have posed questions on the efficiency and effectiveness of these processes that have become more relevant. 

In this paper, we present two empirical studies from the field of requirements engineering. While it serves as one sample topic for a broader problem, Requirements Dependency Classification (RDC) has been a topic of interest for both researchers and practitioners. In particular, we study a fine-tuned \textit{BERT} (Bidirectional Encoder Representations from Transformers) \cite{devlin2018bert}, a recent technique published by researchers from Google, with Random Forest for solving RDC. BERT uses bidirectional training of transformer, a popular attention model, to language modelling, which claims to be state-of-the-art for NLP tasks. We compare BERT with Random Forest (RF), a widely used ML technique that serves as a baseline for comparison. 
%In this study, we explore the question, \enquote{How does fine-tune BERT compare with traditional algorithms on an economical scale?} by comparing models' effectiveness with incurred ROI. 

The objective of this study is to present an alternative method to evaluate ML algorithms. In that sense, we demonstrate the perspective of the returns ML algorithms would generate for the investment done while choosing a particular method for a given problem. 
Our research contributions are as follows
\begin{itemize}
    \item Describe an ML process model for ML classification and perform related ROI modeling. 
    \item Empirically evaluate Random  Forest and fine-tuned BERT for textual classification in the context of requirement dependency classification (RDC) using accuracy and ROI.
    %\item Demonstrate that ROI of ML should be beyond accuracy measure, rather it should transcend to measuring overall effort involved in ML process
\end{itemize}

The remainder of the paper is structured as follows: Section \ref{motivating} provides a motivating example of this study, followed by the description of related work in Section \ref{relatedwork}. Section \ref{RDE} explains requirement dependency, its extraction, practical relevance, and research questions. Section \ref{ROIofML} elaborates our ROI modeling of the ML process. Data used in this study are detailed in Section \ref{data} followed by empirical results in Section \ref{results}. The discussion Section \ref{discussion} details implications and limitations of this study before summarizing conclusions in Section \ref{conclusions}. 

\section{Motivating Example}
\label{motivating}
\begin{figure}[t]
\centering
\includegraphics[scale=0.3]{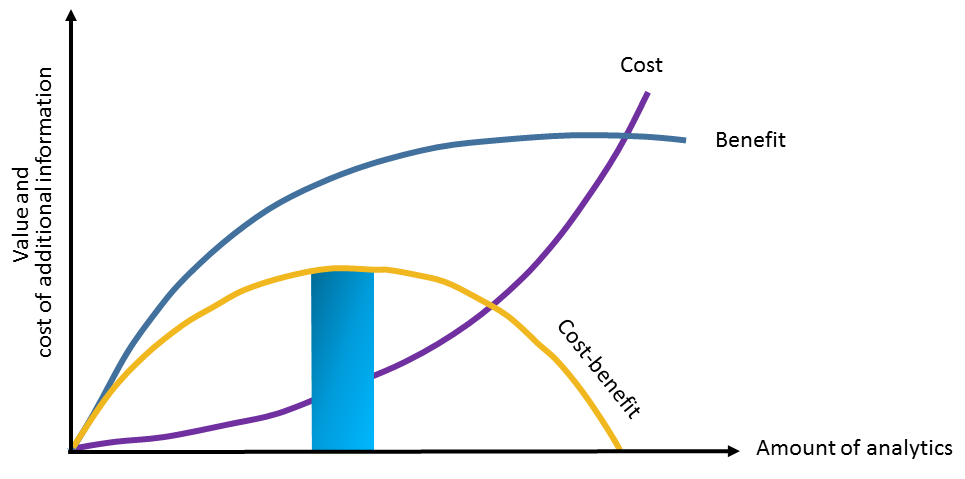}
\caption{Break-even point from cost-benefit analysis of technology investment.}
\label{fig:2}       % Give a unique label
\end{figure}

Figure \ref{fig:2} shows a prototypical ROI curve for technology investment \cite{ruhe2020optimization}. When trying to achieve better results, the investment's cost (or effort) is growing over time, typically non-linearly. However, the benefit achieved from that investment eventually reaches some saturation point beyond which almost no further improvement is achieved. In total, a saturation point is achieved, after which further investment does not pay off anymore (i.e., point of diminishing returns). 

Thus, the most crucial question arises-\textit{Do similar arguments apply for ML classification in Software Engineering?} While this could be true in general, we study it in the context of the requirements dependency classification problem. 

Deshpande et al. \cite{deshpande_ruhe} report the results of a recent survey for requirements dependency classification and maintenance, with 76\% of responses (out of 70) from practitioners. More than 80\% of the participants agreed or strongly agreed that dependency type classification is difficult in practice; dependency information has implications for maintenance, and ignoring dependencies has a significant impact on project success \cite{deshpande_ruhe}.

Applying the advanced NLP technique BERT, we performed an ROI analysis on the requirements dependency classification. Automating this process saves time, and making the classification more effective helps better align the development process with the existing dependencies. For example, if a requirement r depends on another requirement s, then the implementation of s should precede implementing r. Violating this logical dependency will not only delay the usage of r but also decrease the effectiveness of testing.

\begin{figure}[H]
    \centering
    \includegraphics[scale=.43]{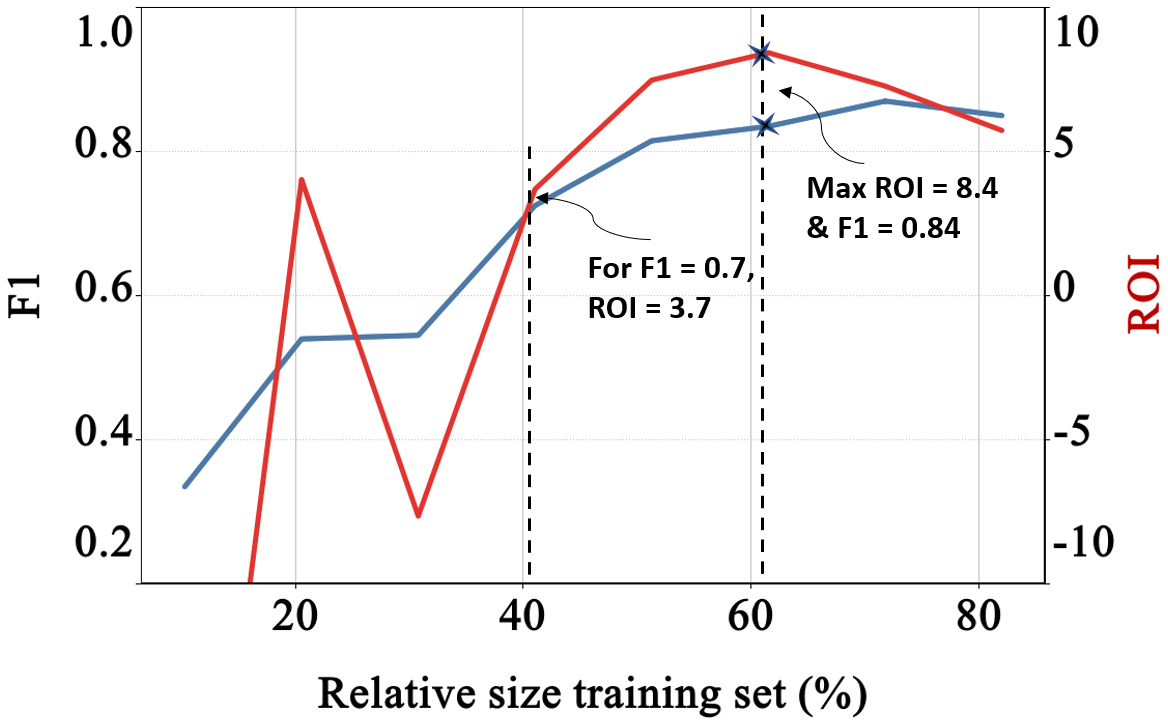}
    \caption{ROI vs F1 of BERT for Firefox dataset \cite{Deshpande_Ruhe_ROI}}
    \label{fig:ESEM}
\end{figure}
Figure \ref{fig:ESEM} shows that there is an early peak in the ROI of using BERT. Since it is a very data-intensive technique, the ROI goes down with increasing training set size before the ROI reaches the global maximum. By comparison, considering only the harmonic mean (F1) of precision and recall gives a different recommendation for training set size. We discuss this in detail in Section 5.

\section{Related Work}
\label{relatedwork}
Although ROI is used in various contexts in Software Engineering and Data analytics, we discuss noted findings from the literature in the context of our proposed research.
% We studied the usage of ROI for decision-making in both Data Analytics and Software Engineering. In addition, we analyze empirical results for RDC.

\subsection{Exploration of ROI in Software Engineering}
Farbey et al. \cite{farbey_evaluation_nodate} explained that as a product moves through its life cycle, various evaluation methods such as ROI, Multi-Objective multi-criteria, Value analysis etc. play an important role in decision making. In this study, ROI was recommended either as a strategy to decrease uncertainty in the business area or to improve knowledge of how technology would operate.

Khoshgoftaar et al. \cite{khoshgoftaar_cost-benefit_2001} presented an interesting case study of a large telecommunication software system and demonstrated a methodology for
cost-benefit analysis of a software quality classification model. The cost and benefit computations were based on the type-I (FP) and type-II (FN) values of classification models. Although these cost-benefit models were ahead of their time, they did not consider the time and effort investment done on data and metrics gathering for cost computation. In another study on calculating ROI in the software product line, Bockle et al. \cite{bockle_calculating_2004} derived cost and benefit estimates based on organization level criteria, such as cost to the organization and cost of reuse. However, this did not involve data analytics of any form.

%Van Solingen \cite{van2004measuring} analyzed the ROI of software process improvement and took a macro perspective to evaluate corporate programs targeting the improvement of organizational maturity. 

The guesswork could be eliminated from the decision-making process while evaluating the profitability of expenditure, which could help measure success over time. For instance, Erdogmus et al.  \cite{erdogmus2004return} analyzed the  ROI  of quality investment to bring its value into perspective; posed an important question, "We generally want to increase a software product's quality because fixing existing software takes valuable time away from developing new software. But how much investment in software quality is desirable?  When should we invest, and where?",  which we think is difficult to quantify yet crucial for the success of software-based products. 

Begel \& Zimmermann \cite{begel2014analyze} gathered and listed a set of 145 questions in a survey of 200 Microsoft developers and testers and termed them relevant for DA at Microsoft. One of the questions: "How important is it to have a software DA team answer this question?", expected answer on a five-point scale (\textit{Essential} to \textit{I don't understand}).  
Although this analysis provides a sneak peek of the development and testing environments of Microsoft, it does not provide emphasis on any form of ROI. Essentially, we speculate that the ROI aspect was softened into asking for the perceived subjective importance through this question.

Boehm et al. \cite{boehm2008roi} \cite{boehm_roi_2004} presented quantitative results on the ROI of Systems Engineering based on the analysis of the 161 software projects in the COCOMO II database. Ruhe and Nayebi \cite{ruhe2016counts} proposed the \textit{Analytics Design Sheet} as a means to sketch the skeleton of the main components of the DA process. The four-quadrant template provides direction to brainstorm candidate DA methods and techniques in response to the problem statement and the available data. In its nature, the sheet is qualitative, while ROI analysis goes further and adds a quantitative perspective for outlining DA.

Ling et al. \cite{ling2005predicting} proposed a system to predict the escalation risk of current defect reports for maximum return on investment (ROI), based on mining historic defect report data from an online repository. ROI was computed by estimating the cost of not correcting an escalated defect (false negative) to be seven times the cost of correcting a non-escalated defect (false positive).

%Ling et al. \cite{ling_maximum_2006} use the algorithm in different test strategies to obtain missing values on test data and to address problems of delayed tests. Using expected total cost, a tree is induced for each test example using altered test costs, whereby test costs are reduced to zero for examples with known values, thus making them a more desirable choice.
\begin{figure*}[!ht]
    \centering
    \includegraphics[scale=.5]{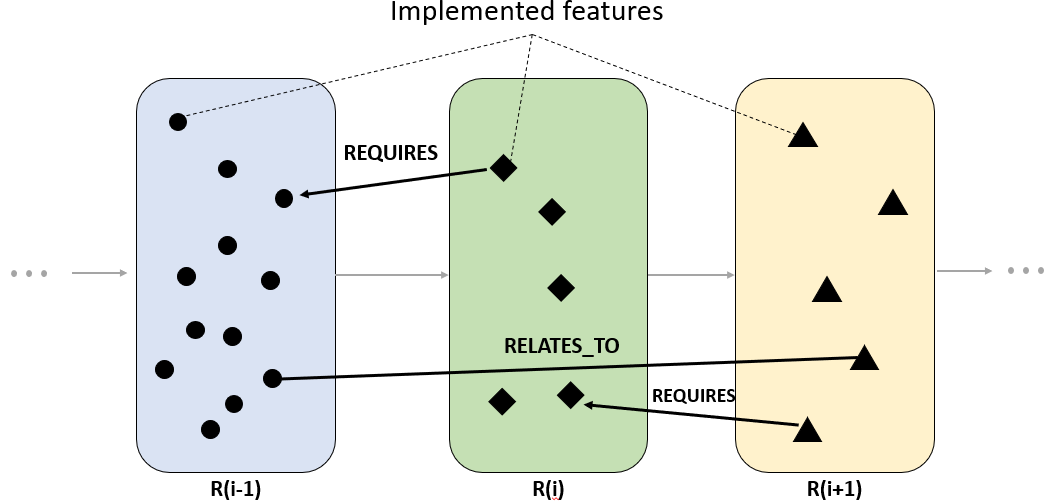}
    \caption{Requirements dependencies across various releases of project}
    \label{fig:RDE}
\end{figure*}
\subsection{Exploration of ROI in Data Analytics}
Ferrari et al. \cite{ferrari2005roi} studied the ROI for text mining and showed that it not only has a tangible impact in terms of ROI but also intangible benefits, which arise from the investment in the knowledge management solution. This solution translates the returns directly that must be considered while integrating the financial perspective of analysis with the non-financial ones.  

Weiss et al. \cite{weiss2008maximizing} emphasized how the quality of external data influence the results and quantified the effort of gathering and using such data when it is available at a premium into cost in terms of CPU time, even though the treatment of the subject is limited to a static setting. In a similar vein, Nagrecha et al. \cite{nagrecha2016quantifying}  proposed a Net Present Value model to determine the cost and impact of analytics programs for an organization.

Taking inspiration from these studies in our research, we not only consider data pre-processing costs as an additional cost aspect but also transform machine learning metrics to dollar amounts, with derived costs and benefits being also validated by industry experts.

\subsection{Empirical Analysis for Requirements Dependency Classification}
Requirements dependencies classification is an active field of SE research. 
The practical importance of the topic was confirmed by a survey \cite{deshpande_ruhe} of over 90 participants from the SE industry. Results showed that more than 80\% of the participants agreed or strongly agreed that (i) dependency type extraction is difficult in practice, (ii) dependency information has implications on maintenance, and (iii) ignoring dependencies has a significant negative impact on project success. 
 
Several empirical studies have explored diverse computational methods that used natural language processing (NLP) \cite{och2002feasibility} \cite{atas2018automated}, semi-supervised technique \cite{deshpande_arora_ruhe_2019}, hybrid techniques \cite{deshpande_ruhe_2020} and deep learning \cite{guo2017semantically} in this context. Recently, Wang et al. \cite{9222252} explored a semi-automatic ML approach based on traceability to identify requirement dependencies to further identify security vulnerabilities. However, none of the approaches considered ROI to decide among techniques and the depth and breadth of their execution level.
\subsection{Exploration of Machine Learning process in Software Engineering}
We analyzed 96 papers from IEEE, Scopus, ScienceDirect, and ACM Digital Library which exclusively used ML, and data analytics within software engineering, and software development domains. Precision, Recall, Accuracy, and AUC were by far the most common performance measures used by researchers in these papers. Additionally, the choice of performance measure was generally not justified. Most studies did not present all steps of the ML process, and most of the papers formally present only 3 steps of the ML process such as data pre-processing, evaluation, and parameter tuning and all these steps are underestimated in terms of effort spent. 

This study highlights the merits of simultaneously considering technical and business criteria when evaluating tradeoffs faced within machine learning approaches for requirements dependency classification (RDC).  We extend prior work that focused on comparing various ML techniques based upon technical criteria of accuracy to include broader consideration of the impact i.e.  Evaluating value generated by the analysis compared to the costs incurred for the analysis.

\section{Requirements Dependency Classification}
\label{RDE}
Similar to requirements elicitation \cite{lim2021data}, extraction of requirements dependencies is a cognitively difficult problem. These dependencies not only influence the development of software but also impact how requirements operate. 
% means that operating on one of the requirements has an impact on operating on the other. 
In this section, we provide the formal problem definition which serves as an example to demonstrate the value of looking beyond accuracy measure and investing in more general concepts of ROI analysis.

\subsection{Practical Relevance}
\par Figure \ref{fig:RDE} is an illustration of the practical relevance of considering requirements dependencies for incremental and iterative software development. Having multiple release cycles: $R_{i-1}, R_i, R_{i+1}$ defines the order of implementing and testing new or updated features. However, if a requirement is implemented in a release $R_i$ but requires a requirement implemented in a later release, then the requirement will not be usable. Similar arguments hold for two requirements that are related to each other but are implemented in different releases. Thus, identifying the dependencies early on is crucial as it drives the implementation as well as testing and rework efforts immensely. 

%In every product development cycle, new and existing requirements having  dependencies occur. Any changes to these dependency types can have a substantial impact on the neighbouring dependencies as depicted in this sub-network. This impact could relate to structural as well as changes in the nature and strength of dependencies.  Additionally, new requirements and change requests can lead to change is dependencies, which could cause conflicts within requirements and increase or decrease the importance of other dependent requirements. Thus, identifying the dependencies early on is crucial as it drives the implementation as well as testing efforts immensely.

\subsection{Problem Formulation}
%For a set of requirements $R$, and any pair of requirements $(r,s)$ $\epsilon$ $RxR$, the symmetric relationship is called a \textit{dependency}, if there is at least one type of dependency ($REQUIRES$, $SIMILAR$, $OR$, $AND$, $XOR$, value synergy, effort synergy) between $r$ and $s$ (independent of type, direction, and strength). 
% Assuming, we can extract basic dependencies, the second research question is to extract the type of the dependency. Details of these types are as following.  

While there are different types of dependencies between requirements \cite{Zhang14}, \cite{Carlshamre} we provide the definitions just for the ones used in the empirical study . For a set of requirements $R$ and a pair of requirements $(r,s)$ $\epsilon$ $R\times R$
\begin{itemize}
%   \item[1)]  An \textbf{\textit{INDEPENDENT}} relationship is defined as the absence of any form of relationship between a pair of requirements.
 
 \item [1)]  Two requirements r, s are called  \textbf{\textit{INDEPENDENT}} if handling one of them has no logical or practical implication for handling the other one. Otherwise, they are called \textbf{\textit{DEPENDENT}}.
 
%\item [1)]  A \textbf{\textit{DEPENDENT}} relationship is defined as the complement set of INDEPENDENT.  i.e., there exists at least one type of the dependency types such as
%\textit{REQUIRES, SIMILAR, OR, AND, XOR, value synergy, synergy} etc. between $r$ and $s$.
 \item [2)] \textbf{\textit{REQUIRES}} is a form of \textit{DEPENDENT} relationship.  If requirement $r$ requires the requirement $s$ to be implemented, then, $r$ and $s$ are in a \textit{REQUIRES} relationship. REQUIRES is an asymmetric relationship.  
\item [3)] \textbf{\textit{RELATES\_TO}} is another specific form of \textit{DEPENDENT} relationship.  Requirement $r$ relates\_to requirement $s$ if changing one of them has an impact on the other. \textit{RELATES\_TO} is a symmetric relationship\footnote{There are other types of dependencies such as \textit{DUPLICATES}, \textit{BLOCKS} etc. that also occur in the these datasets, however, we have considered the ones that occur most frequently}.
% \item [6)] \textbf{\textit{DUPLICATES}} is a special form of \textit{DEPENDENT} relationship.  If $r$ duplicates $s$, or $s$ duplicated\_by $r$, then, $r$ and $s$ are in a \textit{DUPLICATES} relationship  
\end{itemize}

\noindent \textbf{Problem: Binary Requirements Dependency Classification (RDC)}

For a given set $R$ of requirements and their textual description, the binary Requirements Dependency Classification problem (RDC) is to decide for a given pair (r,s) $\epsilon$ 
$R\times R$ if (r,s) is in a \textit{REQUIRES} (called problem RDC\_1) or in a \textit{RELATES\_TO} (called problem RDC\_2)  relationship.

\subsection{Research Questions}
In this paper, two research questions (RQs) are addressed: 

\begin{enumerate}
\item []\textbf{RQ1:} How to model the ROI for ML classification? Specifically, how to instantiate the model for the problem of RDC?\\
\textbf{Rational:} The exclusive consideration of accuracy in the selection of ML classification techniques might be misleading. We consider ROI as an alternative and additional criterion. To study the cost and benefit of the ML classification in a specific context, it is essential to consider the complete process of ML classification and the impact of the results in the original problem space.
\newline
     \item []\textbf{RQ2:} For RDC, how is the preference decision between RDC-BERT and RF impacted by the accuracy criteria F1 that includes ROI?\\
\textbf{Rational:} We evaluate the impact of the selection criteria through two empirical studies on two open-source software (OSS) datasets: Firefox, a software application from Mozilla family \cite{mozillawiki} and Typo3 \cite{typo3}, a content management software. Our goal is to evaluate two extraction techniques (RDC-BERT and RF) to demonstrate the impact of the consideration of ROI in addition to accuracy considerations. 
%    \item [\textbf{RQ2:}] How to model ROI for ML classification, and what is its outlook for RDE?
%\item [\textbf{RQ3:}] For RDE, how does the combined consideration of F1 and ROI influence the method selection (BERT vs RF)?   
\end{enumerate}

\section{ROI Modeling of ML Classification - RQ1}
\label{ROIofML}
Machine Learning classification is an iterative process comprising a series of steps. Aiming at ROI analysis of ML classification requires a look at the effort consumed for all these steps. In what follows, we describe various ML process steps, we estimate cost and benefit, and project the ROI of ML classification.

Although various ML workflow has been defined in the literature \cite{fayyad1996kdd} \cite{amershi2019software} \cite{mlworkflow}, in this section, we present the simplified version of it mainly focusing on the ML process.

% \subsection{ROI of Machine Learning Classification - RQ1}
% \subsection{Modeling ROI for ML classification-RQ2}
% \subsection{ROI modeling}
% The classification algorithms such as RF and NB, have been explored in NLP based SE problems. These algorithms are driven by the feature extraction aspect to a great extent. Thus, could influence their effectiveness on classification outcomes. 
% However, feature extraction is problem specific and incurs substantial cost and access to domain expertise.
% \par On the other hand, BERT eliminates the need for feature extraction since it is a language model based on deep learning. BERT, pre-trained on a large text corpus, can be fine-tuned on specific tasks by providing only a small amount
% of domain-specific data. 

% \section{Modeling ROI Analysis of ML Classification}
\begin{figure*}[!ht]
    \centering
    \includegraphics[scale=.5]{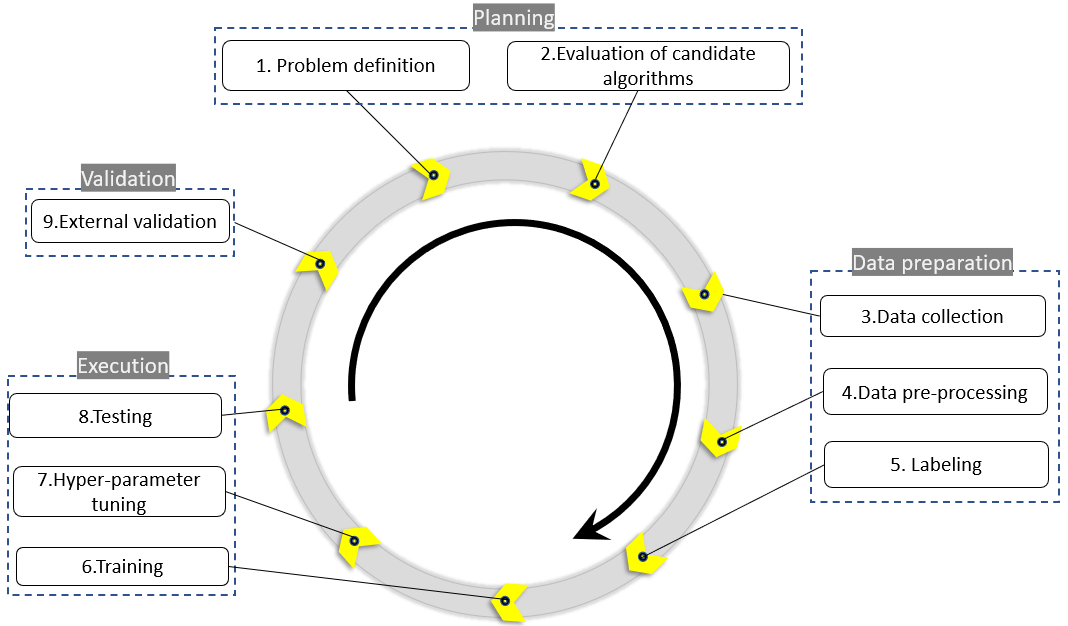}
    \caption{Overview of the steps constituting the ML process}
    \label{fig:arch}
\end{figure*}

\subsection{Modeling the Process}
 The process steps are organized into four Phases: A, B, C, and D called Planning, Data Preparation, Execution, and Validation, respectively. Depending on the context, the effort allocated for these steps may vary. However, this approach parallels the process steps and guidelines for pragmatic optimization in software engineering by Ruhe et al. \cite{ruhe2020optimization}.

An overview of the steps is illustrated in Figure \ref{fig:arch}. Here, we did
not show all the possible arrows to indicate that loops
can, and do, occur between any two steps in the
process. The iterative and interactive ML process, involving various phases is summarized as:
% While the steps are arranged in sequential order, in practice, different types of feedback may occur, and the process is mostly done in different iterations.
\newline \newline
\noindent
\textbf{Phase A : Planning}
\begin{itemize}
    \item[] \textbf{Step 1: Scoping and problem formulation}\\
    Scoping defines the problem context and its boundaries. Problem formulation addresses the key independent and dependent attributes to be considered. As a result of later steps, the problem formulation eventually needs to be adjusted as asking the right question constitutes the largest effort for any application effort. 
%   \item [\textbf {Step 2:}] \textbf{Feature selection}
    %There is the goal to keep the classification model simple, but powerful. Feature selection addresses the step to decide which attributes (features) should be included and which ones could and should be left out. Different algorithms exist to solve this problem \cite{cai2018feature}
    \newline
    \item [] \textbf{Step 2: Evaluation of candidate machine learners}\\
    A variety of ML algorithms exist and new ones are discovered regularly. Commonly used machine learning algorithms include Linear Regression, Logistic Regression, Decision Trees, K-means, Support Vector Machines, Naïve Bayes, Random Forest, and Neural Networks. There is no obvious preference in the sense that "One size fits all". However, there could be recommendations for a particular ML algorithm for a given problem based on its exemplary performance for a similar problem(s). An initial evaluation helps to select the most promising one(s). The selection is influenced by the success criterion of the classification (e.g., accuracy).\\
    \end{itemize}
    
\noindent    
\textbf{Phase B: Data Preparation}
\begin{enumerate}
    \item []\textbf{Step 3: Data collection}\\
    Different sources of data might exist for performing ML classification. Data collection looks into what is potentially relevant and checks the type and availability of the data.
    \newline
    \item []\textbf{Step 4: Data pre-processing}\\
    Raw data would not be ready for processing through the ML algorithm as it could have duplicates, missing values, and contradictions that need to be tackled first for error-free results. Performing such pre-processing operations, for example, data
cleaning, normalization, transformation, feature extraction and
selection, etc. are essential for the success of ML classification, but these steps consume a considerable amount of human resources and processing time. The outcome of data pre-processing is the training
set which could be processed through ML algorithm further \cite{kotsiantis2006data}.
\newline
\item []\textbf{Step 5: Labeling}\\
Labeling is to assign labels to ground truth data \cite{amershi2019software}. Supervised ML methods need labeled data unlike unsupervised ML methods. Labeling is generally performed by domain experts who identify a set of samples (that are most likely representative of the real-world data) to train the ML models. Depending on the nature of the problem, online crowdsourced platforms could also be used for labeling tasks \cite{nayebi2017MAPFEAT}.
\end{enumerate}

\noindent \textbf{Phase C: Execution}
\begin{enumerate}
    \item [] \textbf{Step 6: Training}\\
    The key idea of ML is to learn from existing data and then apply the resulting model to new data. The quality and quantity of the training data are often as important as the actual machine learning algorithm. To learn from existing data also means that the data set is complete, with known input and output of the observations. 
    \newline
    \item [] \textbf{Step 7: Hyper-parameter tuning}\\
    ML algorithms depend upon several parameters such as named model parameters and named hyper-parameters.  Named model parameters can be initialized and updated through the data learning process (e.g., the weights of neurons in neural networks). Named hyper-parameters cannot be directly estimated from data learning and should be set before training an ML model because they define the model architecture. Tuning these parameters means achieving settings that enable good algorithmic performance \cite{kuhn2013applied}.
    \newline
    \item [] \textbf{Step 8: Testing}\\
    After training, the model is applied to the selected test set(s) (a small part of labeled data that is held out and excluded from the training process). The larger the number of variables in the real world, the bigger the training and test data should be. From performing testing, classification error counts are captured in the form of a confusion matrix.
\end{enumerate}

\noindent
\textbf{Phase D: Validation}
\begin{enumerate}
    
    %\item [\textbf {Step 9:}] \textbf{Internal validation}
    %The size and selection of test and training sets influence the results. Internal validation means to vary the settings and synthesize the results. K-fold cross validation is the widely used technique for that \cite{james2013introduction}. 
    \item [] \textbf{Step 9: External validation}\\
    Success from Step 8 does not automatically imply the success of the results in the context of the application. The validity of the problem formulation and the data might prevent the applicability of the results (i.e., not actionable within the organization resulting in significant wasted effort). 
    
    Internal validation approaches such as cross-validation can not guarantee the quality of a machine learning model due to potentially biased training data. External validation is critical for evaluating the generalization ability of the machine learning model, where independently derived datasets (external) are leveraged as validation datasets. While such independent validation is also sometimes used to refer to a validation study by other researchers that the researchers who developed the model \cite{ho2020extensions}. 
\end{enumerate}
% \end{itemize}

%Data processing is an umbrella term used for various tasks that are precursor to most of the  analysis processes \cite{sarker2021machine}. Data processing typically consists series of steps such as data gathering, cleaning, pre-processing to derive useful data etc. Data processing may be an important step of the research process in order to meet statistical assumptions for analytic techniques.

%WE NEED TO PRESENT BASIC CONCEPTS HERE OF THE PROCESS AND THE ROI Measurement OF ML.

%https://medium.com/dataseries/7-steps-to-machine-learning-how-to-prepare-for-an-automated-future-78c7918cb35d

\subsection{Modeling Cost and Benefit}
Acknowledging that ML classification is a process of steps with possibly multiple iterations suggests the need to look at the estimated cost for all these steps. Cost estimation is known to be inherently difficult in software engineering \cite{shepperd2014cost}. The same is true for value prediction. Despite many factors influencing the costs and benefits of ML classifications, we provide a preliminary model to allow a rough estimate of the ROI. 

For cost estimation, we make the assumption that the total cost of performing ML classification with any given ML technique is the sum of cost components of the four phases outlined in the previous section. To simplify the model, we focus on Phase B (Data Preparation) and Phase C (Execution) and ignore the other two phases. Finally, we assume an 80:20 effort (and cost) ratio between Phase B and Phase C, emphasizing the fact that the majority of effort is spent on data preparation. 

%ata processing is an umbrella term used to combine data collection ($C_{dg}$), pre-processing ($C_{pp}$) and labeling ($C_{l}$) under one hood, each one of which is a cost component. However, not all costs are fixed and some vary based on the solution approach used to tackle any decision problem. For example, supervised Machine Learning (ML) requires a large amount of annotated data, to begin with, whereas Active Learning acquires these annotations over a period of time in iterations until a stopping condition for classification operation is reached \cite{settles2009active}.  Additionally, there is a cost associated with modeling and evaluation ($C_{e}$). 

% \textcolor{red} {GOURI: Please provide a model to estimate cost for Phase B, which is similar to what you already did in the ESEM paper.}\\

For modeling the benefit of the classification results, we are looking at classification errors and their cost (penalty) created. A \textit{confusion matrix} CM is a matrix that contains information relating actual with predicted classifications. For \textit{n} classes, CM will be an $n \times n$ matrix
associated with a classifier. Table \ref{CMSample} shows the principal entries of CM for binary classification.
\begin{table}[!ht]
\centering
\renewcommand{\arraystretch}{1.5}
\caption{A confusion matrix of binary (two) class classification problem}
\begin{tabular}{p{1.5cm} |p{2cm}|p{2cm}|}
\cline{2-3}
                         & \textbf{Predicted Negative} & \textbf{Predicted Positive} \\ \cline{2-3}
\multicolumn{1}{l|}{\textbf{Actual Negative}} & True Negative (TN)          & False Positive (FP)         \\ \hline
\multicolumn{1}{l|}{\textbf{Actual Positive}} & False Negative (FN)         & True Positive (TP)          \\ \cline{2-3}
\end{tabular}
\label{CMSample}
\end{table}

\begin{table*}[!ht]
\centering
\caption{Parameters used for ROI computation}
\renewcommand{\arraystretch}{1.3}
\begin{tabular}{l|l|l p{7cm} l}

                                       \multicolumn{2}{l}{} & \textbf{Symbol}   & \textbf{Meaning}                                   & \textbf{Unit} \\ \hline 
\multirow{7}{*}{\textbf{\textit{Cost factors}}{$^1$}} & 
\multirow{1}{*}{\textbf{Phase A}} & $C_{pl}$$^2$           & Planning phase cost                                & \$  \\ \cline{2-5} &
\multirow{2}{*}{\textbf{Phase B}} & $C_{dg}$          & Data gathering cost                                & \$            \\ 
                                       &                                   & $C_{pp}$          & Pre-processing cost                                & \$            \\                     & \multirow{3}{*}{\textbf{Phase C}} & $C_l$             & Labeling cost                                      & \$            \\ \cline{2-5} 
                   
                                       &                                   & $C_t$$^2$             & Hyper-parameter tuning cost                        & \$            \\                 &
                                       & $C_{train/test}$             & Training and testing cost                        & \$            \\ \cline{2-5} 
                                       &
                                       \textbf{Phase D}                  & $C_e$$^2$             & External Validation cost                                    & \$            \\ \hline \hline 
\multicolumn{2}{l|}{\multirow{2}{*}{\textbf{Classification Penalty}}}     & $Cost_{FP}$    & Penalty per FP                                     & \$            \\ 
\multicolumn{2}{l|}{}                                                     & $Cost_{FN}$    & Penalty per FN                                     & \$            \\ \hline \hline
\multicolumn{2}{l|}{\multirow{6}{*}{\textbf{Others}}}                     & $N_{HR}$          & \#Human resources                                  & Number        \\ 
\multicolumn{2}{l|}{}                                                     & $C_{HR}$          & Human Resource cost                                & \$/hr         \\ 
\multicolumn{2}{l|}{}                                                     & $N_{train}$       & Size of the training set                           & Number        \\ 
\multicolumn{2}{l|}{}                                                     & $N_{test}$        & Size of the test set                               & Number        \\ 
\multicolumn{2}{l|}{}                                                     & $N$               & $N_{train}$ + $N_{test}$                           & Number        \\ 
\multicolumn{2}{l|}{}                                                     & $Value_{prod}$$^3$ & Estimated value of the product for a release cycle & \$            \\ \hline
\end{tabular}
\label{Params}
\begin{flushleft}
\textbf{$^1$}\footnotesize{These are per sample cost factors. All the costs are computed by translating them from minutes to \$ by multiplying with resources and cost per hour of the resources} \\
\textbf{$^2$}\footnotesize{For simplicity few of the cost factors have been assumed to be zero}
\\
\textbf{$^3$}\footnotesize{This value was computed using various cost estimates for a period of one release cycle (= 18 months)}
\end{flushleft}
\end{table*}

The F1 score is a measure of the model's accuracy based on the training set and defined as the harmonic mean of the model's precision and recall in (\ref{eq:1}).
\begin{equation} \label{eq:1}
\centering
F1 = \frac{2 \times TP}{2 \times TP + FP + FN}
\end{equation} 

% In the rest of this paper, Random Forest ML and \textit{baseline} are used interchangeablyis referred to as a \textit{baseline}. Specifically, overriding the exclusive accuracy considerations, we compare F1 and ROI from RF and RDC-BERT, and analyze different results depending on the preference criterion.

% Figure \ref{fig:ResearchFlow} shows the overview of this study and premises of RQs. 
% \begin{figure}[ht]
%     \centering
%     \includegraphics[scale=.5]{Images/overview1.png}
%     \caption{Overview of this study}
%     \label{fig:ResearchFlow}
% \end{figure}

In the context of dependency classification, the benefit could be modeled in terms of the ability of the ML model to produce the least amount of overhead by 1) Incorrectly classifying independency as a dependency (False Positive) 2) Incorrectly classifying dependency as independent (False Negative). So, using $Cost_{FP}$ and $Cost_{FN}$ as estimated re-work costs due to classification overhead, $Sum(Cost_{FN}, Cost_{FP})$ would be the cumulative expense that a company has to bear. 
\par In a release cycle, if 
\textit{estimated value} that a product could generate is :$Value_{prod}$ then the $Benefit$ would be the difference of the \textit{estimated value} and the \textit{classification overhead}. Table \ref{Params} lists the relevant cost components and their corresponding units. 

\subsection{Modeling ROI}
During every classification, $Cost$ and $Benefit$ were computed using the parameters explained in Table \ref{Params}. \textit{Cost factors} are data processing costs (Phase B and Phase C) for all the train ($N_{train}$) and test  ($N_{test}$) samples ($n$) in every iteration. This is further translated into dollar-cost by multiplying with hourly charges ($C_{HR}$) of $N_{HR}$ human resources. 
\begin{equation} \label{eq:2}
    % Cost = n *  \frac {(C_{dg} + C_{pp} +  C_e + C_l)}{60}*N_{HR}* C_{HR}
    Cost = n\times\sum_{all\hspace{1mm} applicable}{Cost\hspace{1mm} factors}\times N_{HR} \times C_{HR}
    \end{equation}
$Return$ computations for RDA, assumes reward ($Cost_{FP}$) for misidentifying the independent requirements (FP) and heavily penalizing ($Cost_{FN}$) instances that were falsely identified as independent (FN).
\begin{equation} \label{eq:4}
\centering
     Total Penalty = FP \times Cost_{FP} + FN \times Cost_{FN}
\end{equation}

\begin{equation} \label{eq:5}
\centering
     Benefit = Value_{prod} - Total Penalty
\end{equation}

Return and investment are context-specific terms, and studying the ROI of Machine Learning classification needs tailoring to the context of the study. To determine the ROI, we follow the simplest form of its calculation relating to the difference between $Benefit$ and $Cost$ to the amount of $Cost$ as shown in (\ref{eq:6}). Both $Benefit$ and $Cost$ are measured as human effort in person-hours.
\begin{equation} \label{eq:6}
\centering
ROI = (Benefit - Cost)/Cost 
\end{equation} 
% Costa et al. \cite{costa2005intraoral} distinguished the “hard ROI” from the “soft ROI”. The former refers to the direct additional revenue generated and cost savings. The latter improved productivity, customer satisfaction, technological leadership, and efficiencies. 
The core investigative focus of our study is to evaluate various conditions under which RDC-BERT (fine-tuned BERT using data specific to requirement dependency extraction) is preferable to the baseline ML method: Random Forrest (RF).

In this empirical analysis, beginning with a small train set, classifiers were created, and then the train set was incremented slowly by a fixed factor to generate new classifiers in every iteration until all the data available for training was exhausted. In every iteration, the classifiers were tested for a small fixed data set to capture the results.

% \subsection{Requirements Dependency Extraction Using Active Learning}
% Random sampling (Passive Learning) randomly selects a training set - referred to as \textit{Baseline} in the rest of the paper. Active Learning selects the most informative instances using various sampling techniques such as MinMargin and LeastConfidence \cite{settles2009active}. We compare \textit{Baseline} with AL using RF as a classifier for this scenario. The analysis was done by adding a few training samples in every iteration concurrently to classify the unlabeled instances.

% \textit{Active Learning} (AL) is a ML method that guides a selection of the instances to be labeled by an oracle (e.g., human domain expert or a program) \cite{settles2009active}. While this mechanism has been proven to positively address the question, \enquote{Can machines learn with fewer labeled training instances if they are allowed to
% ask questions?}, through this exploration, we try to answer the question,\enquote{Can machines learn more economically if they are allowed to ask questions?} \cite{settles2011theories}.
% \subsection{Requirements Dependency Extraction Using BERT} 
% We compare two supervised classification algorithms: Naive Bayes (NB) and Random Forest (RF) - ML algorithms successfully and prominently used for text classification\cite{manning2010introduction} in the past, with a fine-tuned BERT model \cite{devlin2018bert}. The analysis was performed for an incrementally growing training set size to capture its impact on F1 accuracy and ROI.

\section{Data and Experiment Setup}
\label{data}

% Various online bug tracking tools are extensively used these days which host all the different types of issues such as tasks, bugs, epics, stories, features, enhancements etc. that would occur during product development. 
Online bug tracking systems such as Bugzilla \cite{bugzilla} and Redmine \cite{redmine} are widely used in open-source software development. Feature requests, tasks, bugs, epics, stories, features, enhancements, and new requirements are logged into these systems in the form issue reports \cite{shi2017understanding} \cite{bhowmik2015resolution} which help software developers to track them for effective implementation \cite{shin2015guidelines}, testing and release planning \cite{ruhe2005art}. 

We mined data from Bugzilla and Redmine related to features for the two OSS projects namely, Firefox -  a Mozilla web browser application and Typo3 - a content management system.
\begin{table*}[!ht]
\centering
\renewcommand{\arraystretch}{1.3}
\caption{Dependency pair samples from the two datasets}
\begin{tabular}{p{2.2cm}|p{1cm} p{5cm}|p{1cm} p{5cm}}
\hline
\textbf{Dependency type} & \textbf{ID} & \textbf{Description}                                                                     & \textbf{ID} & \textbf{Description}                                                                                   \\ \hline
\multirow{8}{*}{\textit{REQUIRES}} & 1432952     & add ability to   associate saved billing address with payment card in add/edit card form & 1429180     & option to use new   billing address when adding new payment card                                                       \\ 
 & 1394451     & update illustration   for error connection failure                                       & 1358293     & ux error connection   failure copy design and illustration update                                                      \\ 
& 1524948     & introduce session   group to allow to manage multiple session at same time               & 1298912     & multiple snapshot   perform periodic session backup and let user restore particular backup    \\ \hline \hline
\multirow{8}{*}{\textit{RELATES\_TO}} & 92822       & ignore   button for link targets                                 & 92297       & make it   possible to mark specific links to not get checked by linkvalidator            \\ 
& 92576       & page   tree filter: make it possible to explicitly filter by uid & 36075       & advanced   filtering for the page-tree                                                  \\
 & 91496       &differentiate   between password reset "by user" and "by admin"  & 89513       &provide   password recovery for backend users                             \\ \hline
\end{tabular}
\label{table:samples}
\end{table*}

\subsection{Firefox}
% Also, data from Bugzilla has been explored for bug report summary utilizing dependency related metadata from bug reports \cite{maalej2017using} \cite{kim2019weighted} \cite{rastkar2014automatic}. Kim et al. \cite{kim2019weighted} utilized \textit{Blocks} and \textit{Depends\_on} association relationships to examine a method for summarizing the bug reports based on weighted-PageRank using these dependency relationships. 
\par In Bugzilla, feature requests are specific types of issues that are typically tagged as “enhancement” \cite{mozillawiki}. We retrieved these feature requests for the Firefox project using the search engine in the Bugzilla issue tracking system and exported all the related fields such as Title, Type, Priority, Product, Depends\_on, and Blocks. Each issue report contains dependency relationships with other issue reports as references metadata \cite{kim2019weighted}.
Using this information, 3,773 depends\_on (also interpreted as \textit{REQUIRES} dependency type) requirements pairs were retrieved. To generate negative samples, requirements that had no relationship were paired and 21,358 samples were generated. 

\subsection{Typo3}
Redmine \cite{redmine} is a free and open-source web-based management and issue tracking tool website. It allows users to manage multiple projects and associated projects. Various issues across a range of projects are updated each day which helps software developers to track them for effective implementation. In Redmine, features are a specific type of issue that is extracted in this paper for further data analysis.
Typo3 Content Management System (CMS) is an Open Source Enterprise Content Management System\cite{typo3} with a large global community of approximately 900 members of the TYPO3 Association. We collected information such as issue\_links, description, the version found, the version released, issue\_id etc. for 5,017 features using Redmine's REST API through a Python script for this study. 
\par All feature descriptions that had fewer than three words in them were filtered out, resulting in 1,324 feature pairs with dependency type \textit{RELATES\_TO}. 
% and 356 pairs with \textit{DUPLICATES}. 
Using the rest of the features that were not in any type of dependency with others, 9,270 pairs were generated as a negative sample set. 

Table \ref{table:samples} mentions sample pairs of requirements dependencies. For example, to be able to associate the address with payment card \textit{REQUIRES} ability to use a new billing address when adding a new payment card.  For both data sets, to perform binary classification, both positive and negative samples are needed for training. Since we only had dependent (positive) samples in the data, we generated negative samples by pairing the requirements which were not related in the given snapshot of the dataset.    
 
 % Please add the following required packages to your document preamble:
% \usepackage[table,xcdraw]{xcolor}
% If you use beamer only pass "xcolor=table" option, i.e. \documentclass[xcolor=table]{beamer}

\subsection{Effort and Value Estimation}
Typo3, currently at released version 11, is a complex content management system that is developed as a hybrid OSS software product. It has a core team of 12 members with varying skills and expertise. They have a major release cycle of 18 months and they plan two or more releases ahead of time. Developers are encouraged to track the dependencies in Jira, however, a few of the team members utilize post-its to work and track them. Typo3 does not explicitly consider Requirements Engineering as a development phase, but they term the efforts towards identifying features and extracting dependencies as conceptual work or scoping. Over 15\% of the release, the cycle is identified as scoping effort and about 25\% of scoping in a release cycle is identified as dependency extraction and identification. Nine team members and the CTO are involved, mostly in identifying the dependencies. 

The CEO confirmed that about 80 \% of the features are in some form of dependency with each other and missing the dependencies is more problematic than misidentifying them. As he puts this in words, \enquote{if you miss dependencies then it starts to ramp up quickly and this is when things go wrong, and breaks deadlines. we wanted to release in April (4 weeks ago) now deadline is mid October}.

Typo3 identifies and manages seven different types of dependencies and their inversions such as precedes, blocks, clones, caused\_by etc. Most of the dependency issues are identified rigorously through testing and the estimated re-work is about 12\%. They have minimal manual testing as they have test suits of over 75,000 test cases. The CEO estimated that the overwork caused by missing dependencies is about 10\% of the efforts. The average salary of the nine people involved in re-work is \$70 (CAD). A summary of all estimates is provided in Table \ref{Assumptions}.
\begin{figure*}[!htpb]
    \centering
    \includegraphics[scale=.5]{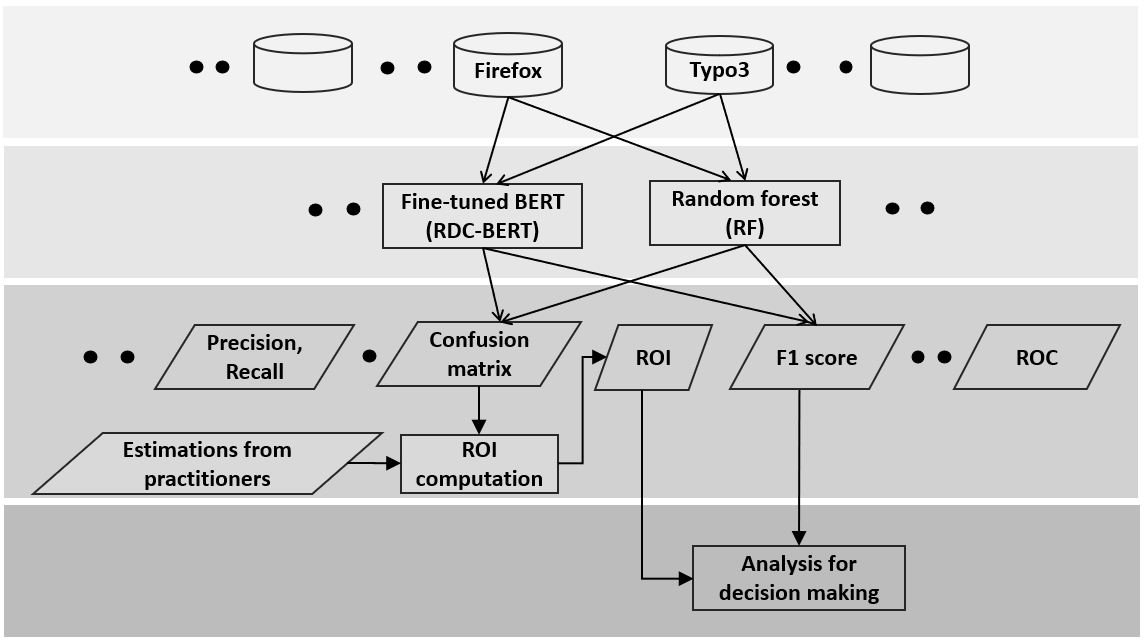}
    \caption{Overview of the experiment setup}
    \label{fig:expSetup}
\end{figure*}
\subsection{Experiment Setup} 
 Figure \ref{fig:expSetup} depicts the overview of our experiment setup. The complete approach is multi layered as highlighted in the shades. Each one of these could be further expanded to include additional elements for solution space evaluation further. 
 
In this study, to generate the results, RF, Naive Bayes and SVM ML algorithms were compared against RDC-BERT for the two datasets: Firefox and Typo3. Overall eight experiments were conducted. Since RF performed better among all the conventional ML algorithms \cite{Deshpande_Ruhe_ROI}, we report the results of RF and RDC-BERT (i.e. totally four experiments). 

For each experiment, we computed ROI using False Negative and False Positive values (from Confusion matrix). In Section \ref{results} we present the insights to aid decision-making in algorithm selection based on these eight outcomes. For additional clarity, we list the names of the analysis of the results and their description in Table \ref{table:eightTest}.

\begin{table*}[h]
\centering
\renewcommand{\arraystretch}{1.25}
\caption{Overview of the various analyses done in Section \ref{results} }
\begin{tabular}{p{1cm} p{4.5cm} p{6.5cm} p{1cm}}
\textbf{} &\textbf{}        & \textbf{Description}                                    &       \\ \hline
Fig \ref{fig:F1_Firefox} &F1\_Firefox                & Firefox: Compare F1 of RF and RDC-BERT & RQ 2.1 \\ 
Fig \ref{fig:F1_typo3} & F1\_Typo3                  & Typo3: Compare F1 of RF and RDC-BERT   & RQ 2.1 \\ 
Fig \ref{fig:ROI_Firefox} & ROI\_Firefox               & Firefox: Compare ROI of RF and RDC-BERT      & RQ 2.1 \\ 
Fig \ref{fig:ROI_typo3} & ROI\_Typo3                 & Typo3: Compare ROI of RF and RDC-BERT    & RQ 2.1 \\ \hline \hline
Fig \ref{fig:BERT_Firefox} & F1\_ROI\_RDC-BERT\_Firefox & Firefox: F1 vs ROI of RDC-BERT              & RQ 2.2 \\ 
Fig \ref{fig:RF_Firefox} & F1\_ROI\_RF\_Firefox       & Firefox: F1 vs ROI of RF                   & RQ 2.2 \\ 
Fig \ref{fig:BERT_Typo3} & F1\_ROI\_RDC-BERT\_Typo3   & Typo3: F1 vs ROI of RDC-BERT                & RQ 2.2 \\
Fig \ref{fig:RF_Typo3} & F1\_ROI\_RF\_Typo3         & Typo3: F1 vs ROI of RF                     & RQ 2.2 \\ \hline
\end{tabular}
\label{table:eightTest}
\end{table*}

Requirements pairs were pre-processed to eliminate noise such as spatial characters and numbers. The generated output is fed to RDC-BERT and RF for training. Care was taken to process the same data snapshot through RF and RDC-BERT models. Further, the fine-tuned BERT model (RDC-BERT) is then used for classification. The data was split (80:20) into train and test sets, and balanced between both classes.

In this empirical analysis, we conducted classification by utilizing a fraction of the whole dataset for training and testing for a small fixed data set. This was repeated by slowly increasing the training set and results were captured.
\newline \newline
\textbf{Random Forest: } For RF, we use TF-IDF to generate word vectors before training.  Also, hyper-parameter tuning was performed and the results for 10-fold cross-validation were computed, followed by testing.
\newline \newline
\textbf{RDC-BERT:} For fine-tuning BERT, a pre-trained BERT model is used in combination with our RDC specific dataset. The result is a fine-tuning BERT model called RDC-BERT. To fine-tune the BERT model, we used \textit{NextSentencePrediction}\footnote{\url{https://huggingface.co/transformers/model_doc/bert.html#bertfornextsentenceprediction}}, a sentence pair classification pre-trained BERT model, and further fine-tuned it for the RDA specific dataset on Tesla K80 GPU on Google Colab\footnote{\url{https://colab.research.google.com/}}.  

In every instance,  for a  given training set size,  RDC-BERT  was trained through three epochs with a  batch size of 32, and a learning rate of 2e-5. In each epoch, the train set was divided into 90\% for training and 10\%for validation. Finally, RDC-BERT was used to classify the test set and the resulting  F1-score and confusion matrix were captured.

% \begin{bclogo}[couleur = blue!10, arrondi=0.1, logo =\bctrombone, barre=none ]{\small{Finding 2}} AL sampling technique out performs PL technique in both Accuracy and ROI\end{bclogo}

BERT eliminates the need for feature extraction since it is a language model based on deep learning. BERT, pre-trained on a large text corpus, can be fine-tuned on specific tasks by providing only a small amount
of domain-specific data. 

% First, for both RF and RDC-BERT, the original data  was  split  into  two  parts  with  a  80:20  ratio.  Thus, 20\% of the complete input data was retained for testing. For  training,  in  the  first  iteration,  5\%  of  the  complete train  set  was  randomly  picked.  Over  the  subsequent iterations,  the  training  set  was  incrementally  increased by  additional  randomly  picked  5\%  and  repeated  until 75\% or more of the training set was exhausted

\section{Empirical Analysis - RQ2} 
\label{results}
In this section, we report the results of our empirical analysis and answer RQ2. We structure results by the type of decisions to be made: (i) \textbf{RQ 2.1: }When comparing two techniques: Which one is preferable under conditions selected?, and (ii) \textbf{RQ 2.2: }When looking at one technique, when to stop the analysis? For both decisions, we present the results of the analysis for the two data sets introduced above and the two techniques under investigation using estimates from Table \ref{Assumptions}.
% \FloatBarrier
\begin{table}[!h]
\centering
\caption{Parameter settings for the two empirical analysis scenarios}
\renewcommand{\arraystretch}{1.25}
\begin{threeparttable}[t]
  \centering
\begin{tabular}{p{4cm} p{3cm}} 
\hline
\textbf{Parameters} &  \textbf{Values}\\ \hline 
Phase B: ($C_{dg} + C_{pp}+C_l$)$^1$  & 1.5 min/sample                    \\   
Phase C: $C_{train/test}$                                 & 0.30 min/sample                       \\  
$C_{HR}$                        &  \$70/hr                   \\  
$N_{HR}$                                   & 10                   \\ \hline
\multirow{2}{*}{$N$} &Firefox:7,546 \\ & Typo3: 2,648 \\\hline 
% $B_{reward}$                          & \$500/TP&          -                \\  
$Cost_{FN}$                        &        \$25,000                      \\  
$Cost_{FP}$                        &        \$10,000                      \\  
$Value_{prod}$             &        \$4,000,000                      \\  
\hline                   
\end{tabular}
\end{threeparttable}
\label{Assumptions}
\begin{flushleft}
\textbf{$^1$}\footnotesize{
$C_{dg}, C_{pp}$ and $C_l$ are weighed equally (= 0.5min/sample) each. Also ratio of Phase B:Phase C = 80:20 has been considered}
\end{flushleft}

\end{table}
\subsection{RQ 2.1: Comparison between RDC-BERT and RF}
\begin{figure*}[!ht]
\begin{minipage}{.45\textwidth}
    \centering
    \includegraphics[scale=.5]{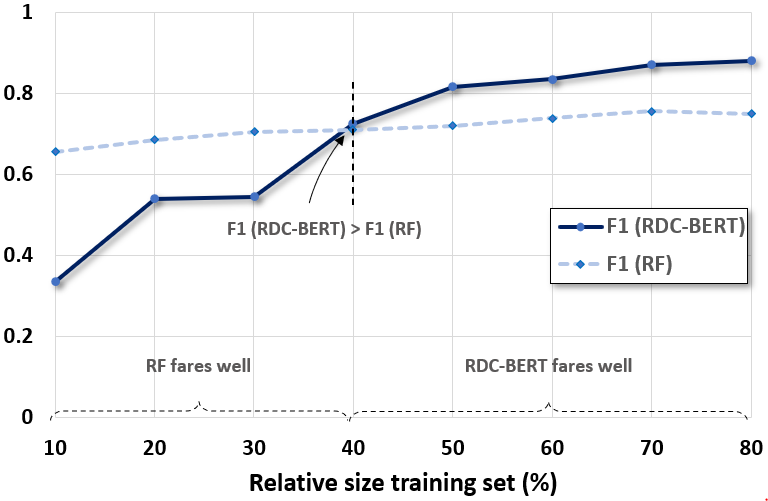}
    \caption{F1 of RDC-BERT vs RF for Firefox dataset}
    \label{fig:F1_Firefox}
    \end{minipage}
    \hspace{10mm}
    \begin{minipage}{.45\textwidth}
    \centering
    \includegraphics[scale=.5]{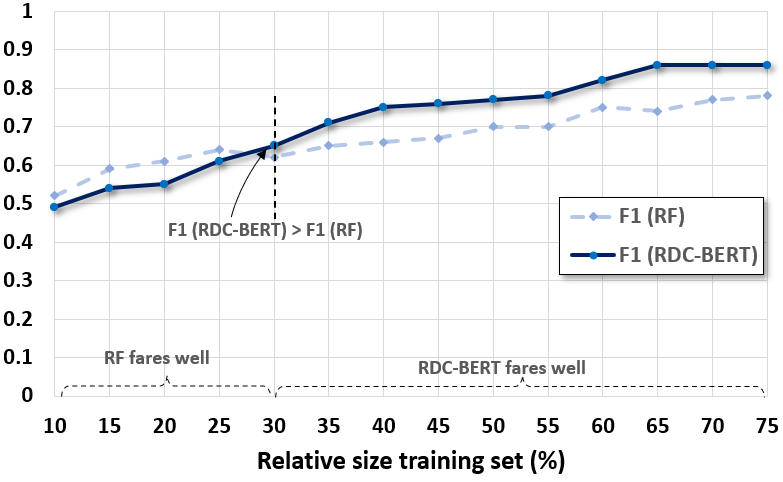}
    \caption{F1 of RDC-BERT vs RF for Typo3 dataset}
    \label{fig:F1_typo3}
    \end{minipage}
\end{figure*}
The traditional approach for comparing techniques is to look at just accuracy for some fixed training set. Figures \ref{fig:F1_Firefox} (F1\_Firefox) and \ref{fig:F1_typo3} (F1\_Typo3) show the comparison of the F1-scores for varying training set sizes for the two datasets. Results show that RF achieves a higher accuracy more quickly for even small-sized train sets respectively. However, with a training set greater than 40\% of the dataset for Firefox and 30\% for Typo3, RDC-BERT achieves better results overall. 

% For Firefox, RDC-BERT started with F1=0.54 and peaked at F1=0.88, showing overall 62\% change in F1-score. Whereas, for RF, beginning with F1=0.7, it reached highest F1=0.75 for 80\% of the dataset, showing only overall 7\% change in F1-score. However, in comparison with RF, RDC-BERT needed at least 40\% or more data to surpass RF's F1-score. 

% For Typo3 dataset, RDC-BERT grew from F1=0.54 to F1=0.86 showing overall 59\% change in F1-score. For RF, beginning with F1=.59 it rose to F1=0.78 showing overall 32\% increase. However, RDC-BERT needed 22\% or more data to surpass RF method. 

\begin{figure*}[!h]
\begin{minipage}{.45\textwidth}
    \centering
    \includegraphics[scale=.45]{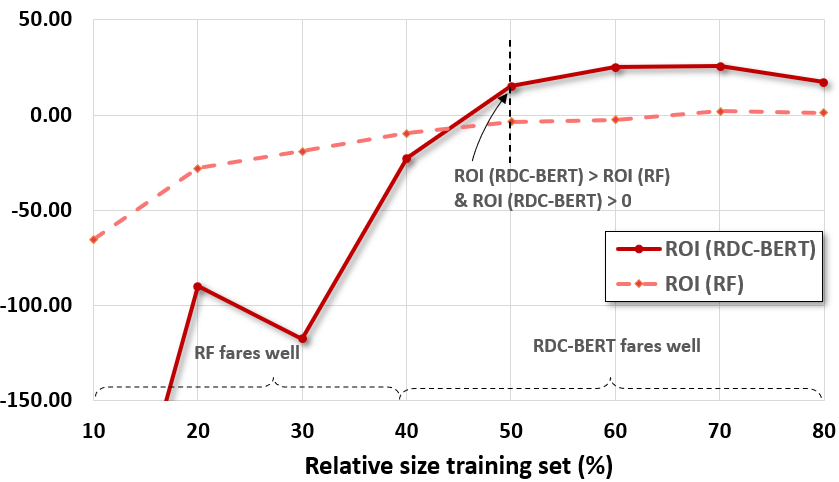}
    \caption{ROI of RDC-BERT vs RF for Firefox dataset}
    \label{fig:ROI_Firefox}
    \end{minipage}
    \hspace{10mm}
    \begin{minipage}{.45\textwidth}
    \centering
    \includegraphics[scale=.5]{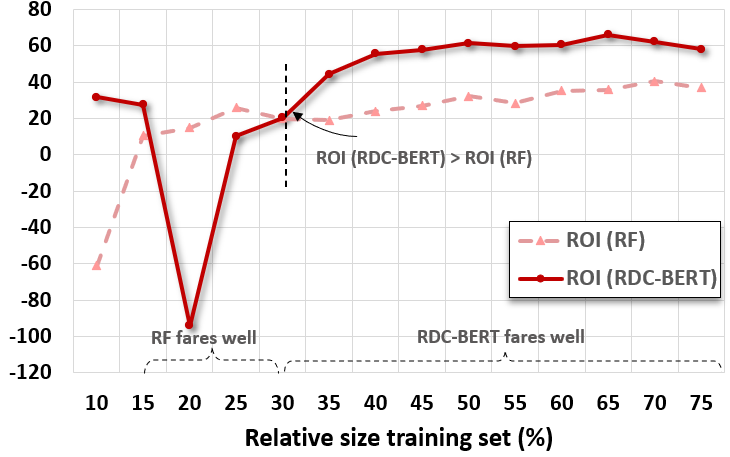}
    \caption{ROI of RDC-BERT vs RF for Typo3 dataset}
    \label{fig:ROI_typo3}
    \end{minipage}
\end{figure*}

Comparison of ROI for the two datasets and two methods (RDC-BERT and RF) is shown in Figures \ref{fig:ROI_Firefox} (ROI\_Firefox) and \ref{fig:ROI_typo3} (ROI\_Typo3) respectively. 
For Firefox, with a smaller-sized train set, RF once again performs better comparatively, even though the ROI is negative. Similar results are evident for Typo3. RF performs marginally better ROI-wise for the smaller training set. ROI of RDC-BERT picks up pace only beyond 40\% and 30\% train set for Firefox and Typo3, respectively.

\subsection{RQ 2.2: Bi-criterion analysis of RDC-BERT and RF}
In the second part of the analysis for RQ2, we look at one technique at a time from the perspective of both F1-score and ROI. This will support decision-making towards the question of when does increase accuracy no longer pays off? 

As illustrated in figures \ref{fig:F1_Firefox} and \ref{fig:F1_typo3}, increased training set does not yield better F1-score beyond 65\%. The F1-score hits a plateau and even starts to degrade for both of the methods and datasets.

\begin{figure*}[!h]
\begin{minipage}{.45\textwidth}
    \centering
    \includegraphics[scale=.47]{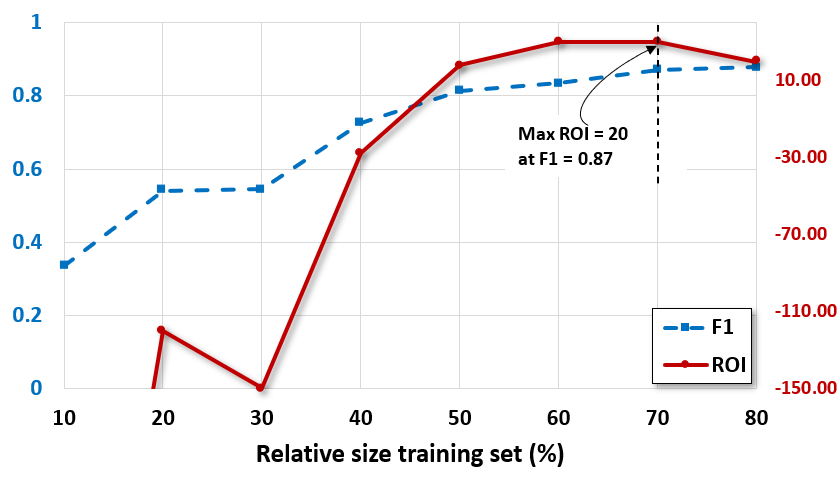}
    \caption{F1 vs ROI of RDC-BERT for Firefox dataset, utilizing values from Table \ref{Assumptions}}
    
    \label{fig:BERT_Firefox}
    \end{minipage}
    \hspace{10mm}
    \begin{minipage}{.45\textwidth}
    \centering
    \includegraphics[scale=.43]{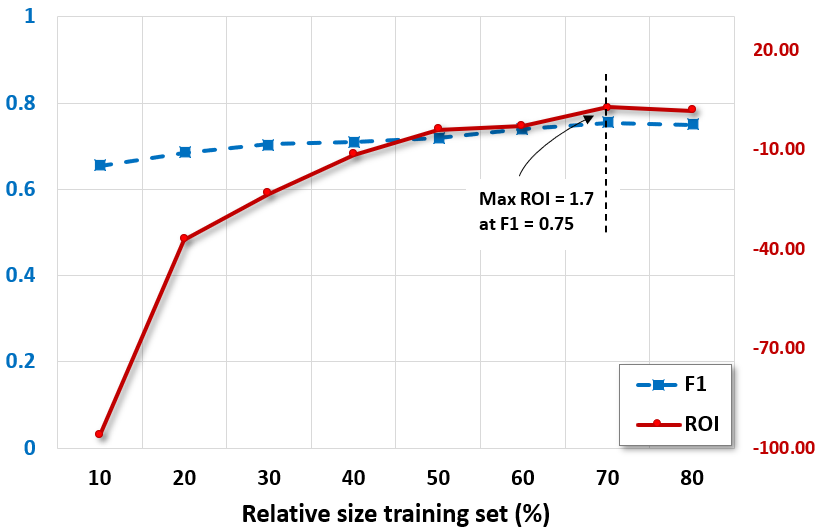}
    \caption{F1 vs ROI of RF for Firefox dataset, utilizing values from Table \ref{Assumptions}}
    \label{fig:RF_Firefox}
    \end{minipage}
\end{figure*}

However, if we look at the trade-off between the F1 and ROI for both datasets, the results become interesting. Figures  \ref{fig:BERT_Firefox}: F1\_ROI\_RDC-BERT\_Firefox show that for RDC-BERT, F1-score increases linearly, however, max ROI is achieved when the train set is 70\% of the dataset. Whereas, for RF, in Figure \ref{fig:RF_Firefox} : F1\_ROI\_RF\_Firefox shows that F1 and ROI for the train set lower than 40\% is better than that of RDC-BERT. Chasing for a higher F1 score does not payoff and one needs to take a closer look at the benefits vs investment in more training data, eventually.

\begin{figure*}[!ht]
\begin{minipage}{.45\textwidth}
    \centering
    \includegraphics[scale=.47]{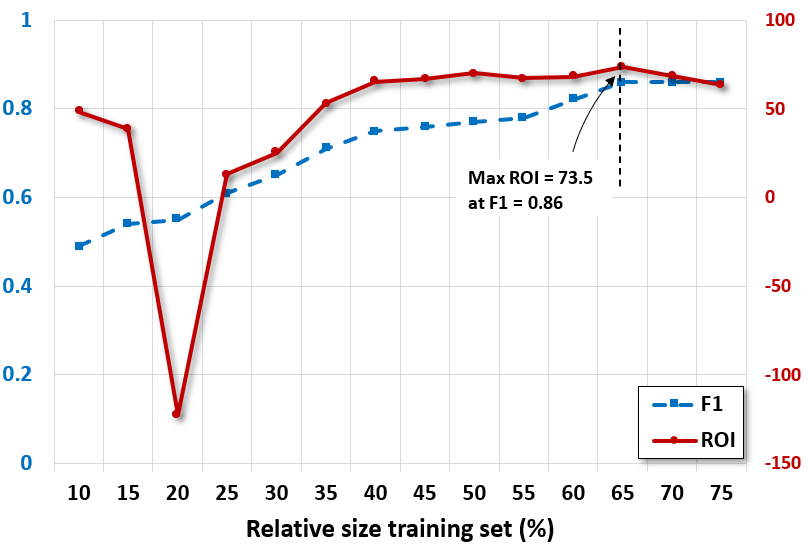}
    \caption{F1 vs ROI of RDC-BERT for Typo3 dataset, utilizing values from Table \ref{Assumptions}}
    \label{fig:BERT_Typo3}
    \end{minipage}
    \hspace{10mm}
    \begin{minipage}{.45\textwidth}
    \centering
    \includegraphics[scale=.5]{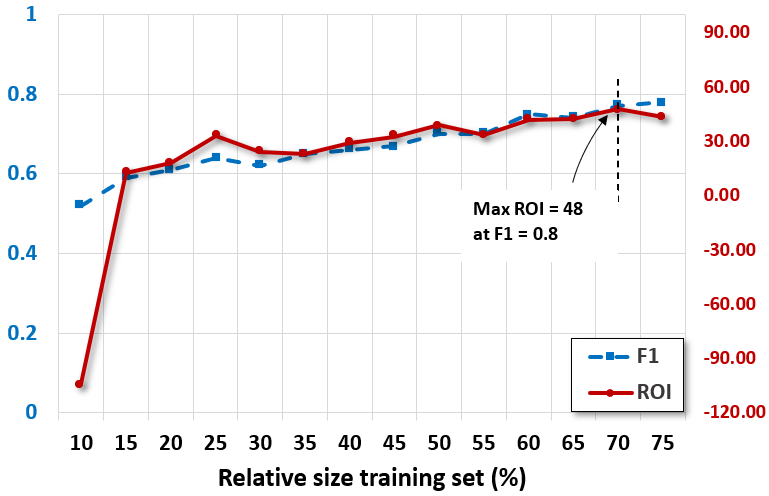}
    \caption{F1 vs ROI of RF for Typo3 dataset, utilizing values from Table \ref{Assumptions}}
    \label{fig:RF_Typo3}
    \end{minipage}
\end{figure*}

For Typo3, in Figure \ref{fig:BERT_Typo3}: F1\_ROI\_RDC-BERT\_Typo3 shows that F1-score and ROI grow steeply for RDC-BERT with the increasing train set. However, similar to Firefox, ROI and F1 of RF are stable and better than RDC-BERT for the train set smaller than 30\%. These findings once again emphasize the need to relook at how F1 and ROI together could aid in deciding on the ML selection.
%\\su\bsecto{Trade-off between ROI and F1-score on ML,  Selection}
% \subsection{BERT vs Random Forest}
%ML algorithms such as random forest and naive bayes, have been explored in NLP based SE problems. These algorithms are driven by the feature extraction aspect to a great extent, which could influence their effectiveness on classification outcomes. However, feature extraction is problem specific and incurs substantial cost and access to domain expertise.

% \subsubsection

%\subsubsection{Firefox}
%For this study, we utilized binary classification. In particular we used \textit{REQUIRES} and \textit{INDEPENDENT} dependency types. In every iteration, dataset size was incremented gradually until all was exhausted and various metrics such as confusion matrix and accuracy were catured. 

%Random Forest algorithm gained a maximum of 2.5 ROI at 0.75 F1 with over 70\% of data used for training (Figure \ref{fig:RF_Firefox}). While accuracy plateaued beyond this point, ROI began to drop. For Fine tuned BERT, at F1=.75, ROI stood at 21 with 50\% of the overall data used for training. Additionally, ROI peaked at 35.8 for F1=0.87 using 70\% data for training. Although accuracy linearly increased with additional data, ROI dropped drastically. 
%\par From this, it is evident that fine tuned BERT performs well comparatively. But to gain a positive ROI both approaches need at least 40\% of training data, which is substantial. With fine tuned BERT it is guaranteed to achieve higher ROI provided there is enough annotated data for training. 
%\subsubsection{Typo3}

%For RF, accuracy and ROI showed similar linear growth beyond 30\% training set. At F1=.8, ROI peaked at 47.7 with 70\% of training set. Although accuracy showed steady increase thereafter, ROI downgraded. In contrast, for fine tuned BERT, ROI showed positive value beyond 20\% training set and achieved ROI = 55.8 at 33\% dataset for F1=.7, evidently outperforming dataset need and accuracy of RF by 10\%. To achieve the peak ROI=78.6 with F1=.86, BERT needed over 66\% of the dataset. Thereafter, ROI started to degrade while F1 hit a plateau.
%\par With just 11\% of the training data (F1 = 0.6), RF starts to generate positive ROI (=18) unlike fine tuned BERT, which needs at least 22\% or more training set (for F1=.6 and ROI=12). However, the fine tuned BERT model pays off the investment in data and drastically outperforms RF. But it is a trade off between investment and likelihood of returns. It is certainly not a bad idea to prefer RF over fine tuned BERT when there is a limitation on data investment and still achieve good results on accuracy.

In both datasets studies, it is evident that RDC-BERT models require large amounts of data (at least 30\% or more) to stabilize and show value (steady positive ROI). When comparing RDC-BERT with RF using ROI criteria (Fig 8 and 9) across the two data sets, RF outperforms RDC-BERT for the lower train set (incurring lower negative returns). However, positive ROIs are observed only at the larger train set at which RDC-BERT is consistently better than RF.  Based upon the Firefox findings (Fig 10 and 11), RDC-BERT approaches the 80\% benchmark accuracy with approximately 50\% of the training data while RF requires 70\% training data to attain the same level of accuracy.  However, both techniques can achieve positive ROI with as little 50\% training data but RBC-BERT achieves maximum ROI (30) with an accuracy of 0.87 with approximately 70\% training data, while RF achieves maximum ROI (2.2) with an accuracy of 0.75 with approximately 70\% training data.  

Based upon the Typo3 findings (Fig 12 and 13), RDC-BERT approaches the 80\% benchmark accuracy with approximately 55\% of the training data while RF requires 70\% training data to reach the same level of accuracy.  However, both RBC-BERT and RF can achieve positive ROI with as little 15\% training data, but RBC-BERT achieves maximum ROI (73.5) with an accuracy of 0.86 with approximately 65\% training data, while RF achieves maximum ROI (48) with an accuracy of 0.80 with approximately 70\% training data.  Thus, RBC-BERT can deliver much higher ROI and similar levels of accuracy than RF given approximately the same amount of training data.  

Finally, the parameter settings that seeded the initial model (Table 5) were based upon industry estimates, which were possible were verified by senior management in the respective firms.  However, some of the findings may be sensitive to these initial conditions.  Thus, these would need to be set for the specific context upon which the data sets are based.  This is also the basis upon which scenario analysis could be conducted to evaluate the worst case, best case and most likely initial conditions to evaluate the impacts on subsequent decisions.

%\begin{bclogo}[couleur = blue!10, arrondi=0.1, logo =\bctrombone, barre=none ]{\small {Finding 2}} Fine tuned BERT model outperforms Random forest when more training dataset is made available. However, it is a tradeoff between investment in gathering more data and highest accuracy that could be achieved with the best possible ROI \end{bclogo}

%Comparison of ROIs for both the methods for the two datasets is as shown in figures \ref{fig:ROI_Firefox} and \ref{fig:ROI_typo3}. For Firefox dataset, RDC-BERT is a clear winner, however, to surpass RF's ROI, it needed 40\% or more data for training. Conversely, for the training data with less tha 40\% of the overall data, RF incurred better ROI comparatively.

%For Typo3 dataset, RDC-BERT is a clear winner as well. However, similar to the Firefox dataset, RDC-BERT needed at least 27\% or more data for training to achieve ROI better than RF. Alternately, we could infer that the RF method fared better with training size of 27\% or less and yet yielded better ROI than RDC-BERT. Interestingly, ROI for Typo3 for both the methods started to degrade beyond 70\% data. 

\section{Discussion}
\label{discussion}
\subsection{Implications}
ML is not simply a cost of doing business, rather it is a foundational  activity that can provide value for the money invested.
Our proposed approach aligns this notion with the strategic direction of the organization.  While return on investment (ROI) is a common approach used for business planning and decision making, it is not applied as widely within software engineering or specifically within applied ML.  

In our study, we demonstrate how to instantiate ROI in the context of RDC. Our approach provides a pragmatic link between the business and technical aspects of the organization by providing a common language that incorporates both the technical aspects inherent in the evaluation of accuracy, with the business considerations of costs and benefits.  We argue that this is an extremely powerful approach that provides evidence that is compelling and consistent for both technical and business decision-making. 

In addition, we think that the ROI approach could sensitize the ML team to the entire process of ML classification and how that process fits into organizational processes. The ROI approach is essential for evaluating the possible tradeoffs between accuracy and the benefits. Mainly because without consideration of the key dependencies within the process, benefits in one part of the process (e.g., improved accuracy) can easily be undermined by excessive costs in another part of the process that would not typically be considered if focused exclusively on accuracy.  Alternatively, lower levels of accuracy in the ML process might be acceptable if other benefits are accruing at reasonable costs.  Thus, valuable ML investments are potentially being avoided based upon not meeting accuracy expectations, when those ML solutions could be sufficient to realize high payoffs for the organization.  

Our approach increases the transparency of the decision-making process by adding diversity to the evaluation criteria that foreground the various tradeoffs being made.  The development of AI tools that businesses and consumers can trust is essential for their continued adoption, especially as there is increasing regulatory scrutiny of the biases that arise in the ML algorithms or inherent in the data used for training.

While ML algorithms are generally trusted for relatively mechanical well-defined problems, this trust plummets when the decisions are subjective, and likely to vary by contextual variables that are not well understood.  This in turn increases the pressure to adjust ML algorithms for variations in specific markets further driving up development costs.  Such pressure directs the focus on customizing products and services based upon ML algorithms for specific markets while increasing costs further and undermines the benefits for certain markets or customers \cite{hbr}.  
The proposed approach considers technical and business aspects simultaneously and provides a more traceable set of interconnected processes.  This approach includes business and technical considerations to enable management to evaluate the risks of some undesirable decisions and the tradeoffs needed to realize the likely benefits. 

\subsection{Limitations}
We have explored RF and RDC-BERT in the context of the RDC problem and presented our results. Since there is no single method which could work for any given problem, comparison of multiple approaches and their results remains out of the scope of this study.

Another threat to validity is the related to the conclusions made. Although we have taken care to randomize the data by shuffling and used stratified split to take care of balanced data in both training and validation, multiple runs with varying first iteration data sample are needed to be more confident on the conclusions made. However, we argue that the key observations made are valid from the restricted empirical validation performed.  

\section{Conclusions and Future Work}
\label{conclusions}
ML classification is widely used in many disciplines of Science and Engineering. In this study, we demonstrate that just looking at performance measures such as accuracy could be misleading when, for example, deciding between two ML techniques evaluated for solving the same problem. Conversely, ignoring the cost and benefit of such a classification could cause the risk of unprecedented emphasis on improving accuracy that might not generate any value for the additional efforts spent. Additionally, in this research, we also provide a high-level ML process for classification (supervised machine learning). However, with minuscule changes, this process can be adapted to unsupervised or semi-supervised ML methods easily.

We use Requirements Dependency Classification as a sandbox to build a proof of the concept based on the two ML techniques used to solve RDC. In the future, we will extend the results in various dimensions. The concepts of this paper will be applied and evaluated for problems from other domains. However, the challenge is to project the benefit of achieving better accuracy results and estimating the total effort of data analysis. Also, depending on the problem, we will investigate other ML techniques and additional data sets.

With a broader data and knowledge base, we aim at developing a customized recommendation system that would support practitioners in their decision-making in terms of "How much Data Analytics is Enough".

%\cite{pfleeger1999albert}
%\cite{shrikanth2020early}

\begin{acknowledgements}
We would like to thank graduate students Saipreetham Chakka and Aris Aristorenas for their assistance in generating results and conducting literature review for this study. \end{acknowledgements}

% Authors must disclose all relationships or interests that 
% could have direct or potential influence or impart bias on 
% the work: 
%
\section*{Conflict of interest}
The authors declare that they have no conflict of interest.

% BibTeX users please use one of
% \bibliographystyle{spbasic}      % basic style, author-year citations
%\bibliographystyle{spmpsci}      % mathematics and physical sciences
%\bibliographystyle{spphys}       % APS-like style for physics
%\bibliography{}   % name your BibTeX data base
\bibliographystyle{IEEEtran}
\bibliography{references.bib}
% Non-BibTeX users please use
% \begin{thebibliography}{}
%
% and use \bibitem to create references. Consult the Instructions
% for authors for reference list style.
%
% \bibitem{RefJ}
% Format for Journal Reference
% Author, Article title, Journal, Volume, page numbers (year)
% % Format for books
% \bibitem{RefB}
% Author, Book title, page numbers. Publisher, place (year)
% etc
% \end{thebibliography}

\end{document}